%% file: main-cikm23.tex
\documentclass[sigconf]{acmart}

\usepackage{todonotes}
\usepackage{mathtools}
\usepackage{subfig}
\usepackage{multicol}
\usepackage{multirow}
\usepackage[ruled, lined, linesnumbered, commentsnumbered, longend]{algorithm2e}
\theoremstyle{definition}
\newtheorem{definition}{Definition}[section]

\SetCommentSty{mycommfont}
\usepackage{algpseudocode}
\usepackage{tcolorbox}
\algnewcommand\algorithmicforeach{\textbf{for each}}
\algdef{S}[FOR]{ForEach}[1]{\algorithmicforeach\ #1\ \algorithmicdo}

\AtBeginDocument{%
  }

\copyrightyear{2023}
\acmYear{2023}
\setcopyright{rightsretained}
\acmConference[CIKM '23] {Proceedings of the 32nd ACM International Conference on Information and Knowledge Management}{October 21--25, 2023}{Birmingham, United Kingdom.}
\acmBooktitle{Proceedings of the 32nd ACM International Conference on Information and Knowledge Management (CIKM '23), October 21--25, 2023, Birmingham, United Kingdom}
\acmPrice{}
\acmISBN{979-8-4007-0124-5/23/10}
\acmDOI{10.1145/3583780.3614869}

\begin{document}

\title[ASTTrans]{Evaluating and Optimizing the Effectiveness of Neural Machine Translation in Supporting Code Retrieval Models: A Study on the CAT Benchmark}


\author{Hung Phan}
\orcid{0000-0001-7464-1597}
\affiliation{%
  \institution{Iowa State University}
  \streetaddress{P.O. Box 1212}
  \city{Ames}
  \state{Iowa}
  \country{USA}
  \postcode{50010}
}
\email{hungphd@iastate.edu}

\author{Ali Jannesari}
\orcid{0000-0001-8672-5317}
\affiliation{%
  \institution{Iowa State University}
  \streetaddress{P.O. Box 1212}
  \city{Ames}
  \state{Iowa}
  \country{USA}
  \postcode{50010}
}
\email{jannesar@iastate.edu}

\renewcommand{\shortauthors}{Phan and Jannesari}

\begin{abstract}
  \input{content/abstract.tex}
\end{abstract}




\maketitle
\input{content/Introduction.tex}
\input{content/Motivation.tex}
\input{content/Background.tex}
\input{content/Definitions.tex}
\input{content/Approach.tex}
\input{content/ExperimentMix.tex}

\input{content/AblationStudy.tex}
\input{content/RelatedWork.tex}
\input{content/ThreatsToValidity.tex}
\input{content/Conclusion.tex}
\input{content/Acknowledgement.tex}

\clearpage
\bibliographystyle{ACM-Reference-Format}
\bibliography{content/refs}
\end{document}

%% file: content/Abstract.tex
Neural Machine Translation (NMT) is widely applied in software engineering tasks. The effectiveness of NMT for code retrieval relies on the ability to learn from the sequence of tokens in the source language to the sequence of tokens in the target language. While NMT performs well in pseudocode-to-code translation \cite{025_SPoC}, it might have challenges in learning to translate from natural language query to source code in newly curated real-world code documentation/ implementation datasets. In this work, we analyze the performance of NMT in natural language-to-code translation in the newly curated CAT benchmark \cite{004_CATBenchmark} that includes the optimized versions of three Java datasets TLCodeSum, CodeSearchNet, Funcom, and a Python dataset PCSD. Our evaluation shows that NMT has low accuracy, measured by CrystalBLEU \cite{009_CrystalBLEU} and Meteor \cite{010_Meteor} metrics in this task. To alleviate the duty of NMT in learning complex representation of source code, we propose ASTTrans Representation, a tailored representation of an Abstract Syntax Tree (AST) using a subset of non-terminal nodes. We show that the classical approach NMT performs significantly better in learning ASTTrans Representation over code tokens with up to 36\% improvement on Meteor score. Moreover, we leverage ASTTrans Representation to conduct combined code search processes from the state-of-the-art code search processes using GraphCodeBERT \cite{002_GCB} and UniXcoder \cite{001_UniXcoder}. Our NMT models of learning ASTTrans Representation can boost the Mean Reciprocal Rank of these state-of-the-art code search processes by up to 3.08\% and improved 23.08\% of queries' results over the CAT benchmark.

%% file: content/Introduction.tex
\section{Introduction}

Although Neural Machine Translation (NMT) has been proven effective in pseudocode-to-code translation \cite{025_SPoC,026_Scaffold}, applying NMT on practical datasets of real-world NL queries and code snippets might fail due to two reasons. First, in real-world benchmarks, the Natural Language (NL) queries are written to summarize long and complex code snippets, as shown in the study of Barone et al. \cite{050_miceli-barone-sennrich-2017-parallel}. NL query is a type of code documentation used to explain how its source code works with short descriptions as summarization. The lack of mapping between each Line of Code (LOC) to its description in code snippets of CAT benchmark might reduce the quality of NMT translation models on datasets in CAT benchmark compared to pseudocode-to-code translation. Second, NMT usually outputs incomplete code snippets that require error localization and fixing \cite{025_SPoC}. It considers the output a sequence of code tokens instead of an Abstract Syntax Tree (AST) representation. Unlike translation, code retrieval by code search \cite{001_UniXcoder} can return complete code. The idea of code search is to consider the input in the form of NL description from developers as a \emph{query} and each source code snippet in a source code dataset as a \emph{candidate}. Embedding models such as UniXcoder \cite{001_UniXcoder} and GraphCodeBERT (GCB) \cite{002_GCB} then learn the representation as vectors for query and candidates. Next, the best candidate for each query is returned by the search process that finds the candidate with the highest similarity measured by its embedding to the query's embedding.

In this work, we analyze the performance of NMT on learning specific information about the source code by AST's subset of non-terminal nodes, compared to learning from NL query to code tokens. We propose ASTTrans, an NMT translation engine trained using OpenNMT toolkit \cite{003_OpenNMT} to learn the mapping from documentation to our tailored representation of an AST by non-terminal nodes. Then, we build a new code search approach that integrates ASTTrans to the state-of-the-art (SOTA) code search process embedding models \cite{001_UniXcoder,002_GCB}. Our experiments show that ASTTrans improves  code search using SOTA approaches GraphCodeBERT and UniXcoder thanks to its augmented code search process.  We use four datasets of the CAT benchmark \cite{004_CATBenchmark}, for our evaluation. Overall, our contributions are as follows:
\begin{enumerate}     
\item We analyze and demonstrate NMT in learning our tailored representation of AST compared to learning the sequence of code tokens.
\item We build a query-to-ASTTrans Representation model and integrate its output to improve the accuracy of the SOTA code search models GraphCodeBERT \cite{002_GCB} and UniXcoder \cite{001_UniXcoder} and achieve up to 3.08\% MRR improvement on TLC dataset and 1.06\% on average on all datasets of CAT benchmark.
\item We analyze how the parameters of ASTTrans can impact the performance in code search.
\item We conduct a case study to investigate the reasons when ASTTrans can or cannot improve the code search for SOTA models.
\end{enumerate}
The rest of this paper is organized as follows. In section 2, Motivation Example, we introduce an example of a query/ candidate for a code search problem. Section 3 provides background information, summarizing the approaches provided in existing embedding tools GraphCodeBERT\cite{002_GCB} and UniXcoder \cite{001_UniXcoder}. Section 4 shows definitions related to our proposed representation  of AST. Section 5 describes in detail our approach to integrating ASTTrans into the original code search process by SOTA approaches. Section 6 mentions our experiments, including configurations, metrics, and results of our proposed research questions. In section 7, we conduct a case study about when ASTTrans can/cannot improve original models. The remaining sections are Related Work, Threats to Validity, and Conclusion. The replication package is available here\footnote{https://github.com/pdhung3012/ASTTrans/}.

%% file: content/Motivation.tex
\section{Motivation Example}
\input{figTabAlgs/fig_motiv_icse}
Figure \ref{asttrans:fig:motivExample} shows a motivation example of a code search process. Users inputs query described in NL. The output for code search's users is the list of candidates as code snippets sorted by their relevancy to the requirement specified by the query. A good code search system tends to return the correct candidate corresponding with a query as the first (called the top-1) candidate of the output list of candidates. In the following example in Figure \ref{asttrans:fig:motivExample}, this query asks how to perform the $add$ function to a $Map$ object in Java. The correct candidate (Candidate 1) accepts the key, the value as a pair to add, and the map as arguments of a method declaration. It performs the check for the validity of the key/value pair before putting it onto the map object. The incorrect candidate (Candidate 2) attempts to put two pairs on a newly constructed map, which doesn't satisfy the requirement provided by the input query. The query, Candidate 1 and Candidate 2, are extracted from the TLCodesum dataset of the CAT benchmark \cite{004_CATBenchmark}. By UniXcoder \cite{001_UniXcoder}, the result of this query returned candidate two as the top-1 candidate of the output list, meaning UniXcoder returned incorrectly for this query.  

\textbf{Non-terminal nodes of AST.} The sub-ASTs of Line 3 of Candidate 1 and Candidate 2 are shown on the right side of Figure  \ref{asttrans:fig:motivExample}. We use AST-treesitter \cite{023_treesitter} for AST generation for these candidates. The sequence of terminal nodes of an AST of a code snippet generated by AST-treesitter is the sequence of code tokens of that snippet. We have two observations from this example. First, the differences between Line 3 of the two candidates are shown at both the terminal and non-terminal levels. In Candidate 1, non-terminal nodes of Line 3 represented an $if$ statement with information about two nodes of type $binary\_expression$ nested in a sub-AST with the root as another node with type $binary\_expression$ (node $C\_1\_6$). In Candidate 2, a node with type $local\_variable\_declaration$ is the ancestor of a $method\_invocation$ (node $C\_2\_9$). Second, while code tokens can be considered a sequence of terminal nodes, we can also represent the ancestors of code tokens by a set of non-terminal nodes. For Candidate 1 and Candidate 2, the sets of parent nodes that can generate all terminal nodes for Line 3 of Candidate 1 and Candidate 2 can be shown in Figure \ref{asttrans:fig:motivExample}. While there are 14 terminal nodes in the sub-AST of Line 3 of Candidate 1 and eight terminal nodes in the sub-AST of Line 3 of Candidate 2, there are only seven nodes as parent nodes of terminal nodes for Candidate 1 and five nodes represented for Candidate 2. 

%% file: figTabAlgs/fig_motiv_icse.tex
\begin{figure*}
    \centering    
    \includegraphics[width=0.8\linewidth]{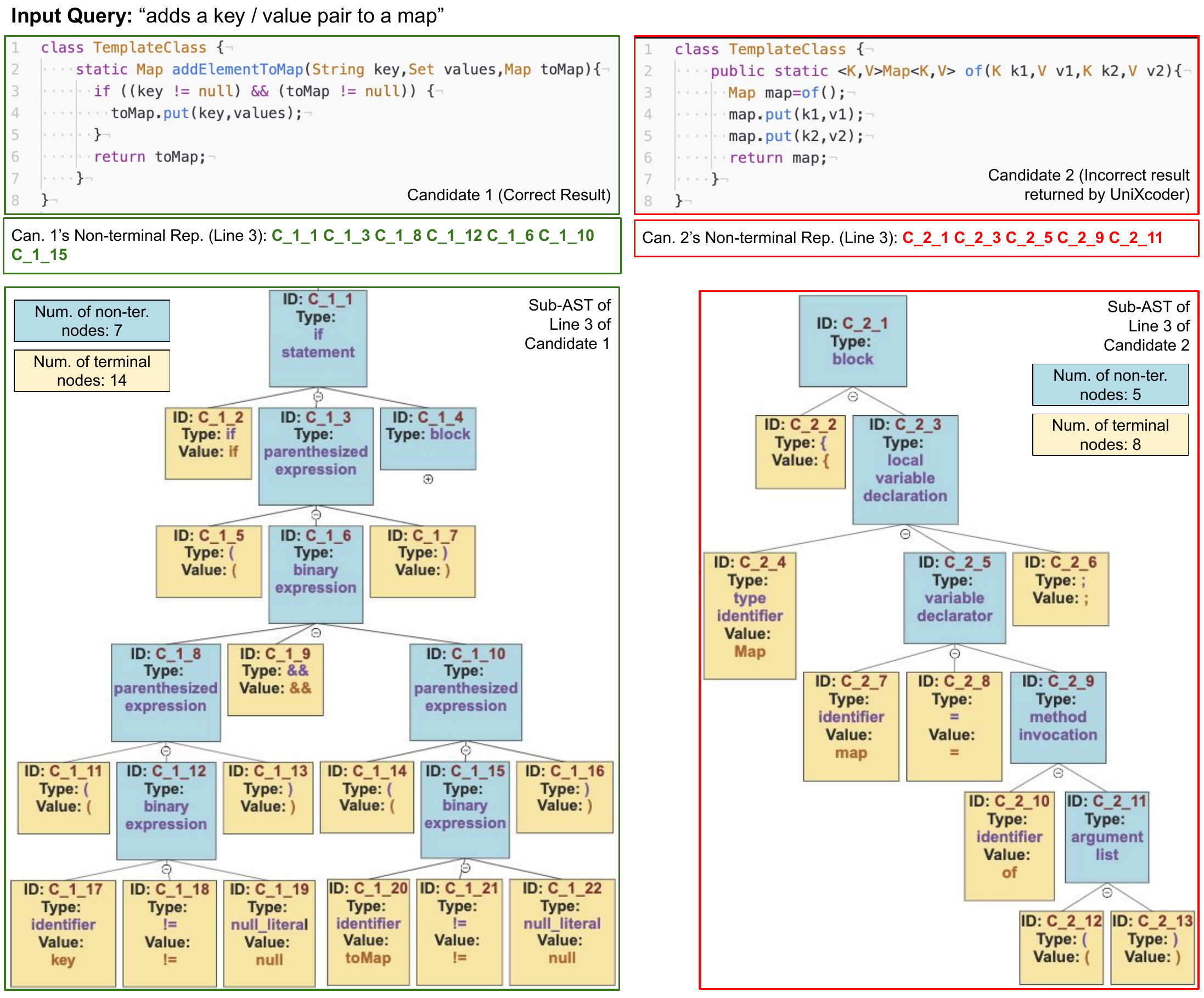}
    \caption{Motivation Example with Input Query, Correct Candidate and Incorrect Candidate Returned by UniXcoder}
    \label{asttrans:fig:motivExample}
\end{figure*}

%% file: content/Background.tex
\section{Background}
\textbf{Natural Language to Code Search (Code Retrieval).} The process of code search by SOTA approaches \cite{002_GCB,001_UniXcoder} is done in two steps (see steps 1 and 2 of Figure \ref{asttrans:fig:overview}). The inputs of code search are a query in NL  and a list of candidates as code snippets. In step 1, the embedding of the query and embeddings of candidates are generated. In step 2, the cosine similarities between the query's and each candidate's embedding are calculated into a matrix of similarities. Based on this matrix, the candidates will be sorted descendingly by the cosine similarity between their embedding and the query's embedding. The output of code search is the list of candidates so that the higher a candidate is ranked in the list, the more relevant to the query it is. The best candidate suggested by the embedding model is the top-1 candidate by the code search process. In these steps, the most important step is generating vectors for queries and candidates.  We use two SOTA models for this step: GraphCodeBERT \cite{002_GCB} and UniXcoder \cite{001_UniXcoder}.

\textbf{GraphCodeBERT (GCB) and UniXcoder.} Applications as downstream tasks by GraphCodeBERT and UniXcoder are built by two processes: building their pre-trained models by pre-training tasks and fine-tuning them. GraphCodeBERT uses a data-flow graph from AST, which highlights the roles of variables to be an input of pre-training tasks. UniXcoder accepts the input as the flattened sequence of all non-terminal and terminal nodes of AST of the source code candidate for pre-training tasks.

%% file: content/Definitions.tex
\section{Sequence-Based Non-terminal Nodes Representation for AST}
This section presents definitions for our tailored representation of AST.  
\begin{definition}[ASTTrans Representation at Depth-K of a Terminal Node]\label{asttrans:def:representationDepthK}
Given a terminal node $t$ inside an AST $a$, the representation at depth-k of $t$, called $DepRep(t, a, k)$, is the ancestor that has its depth (to the root of $a$) equal to $k$ if the depth of $t$ is greater than $k$, or the parent node of $t$ if the depth of $t$ is less than or equal to $k$. We can formulate this function by Formula \ref{asttrans:eq:asttransDepRep}:
\begin{equation}
    DepRep(t,a,k)=\left\{\begin{matrix}
ancestors(t,a)[k] \, if \, k <=length(ancestors(t,a)) 
\\ 
parent  \, node \,  of \,  t  \,  if \, k >length(ancestors(t,a)) 
\end{matrix}\right.
    \label{asttrans:eq:asttransDepRep}
\end{equation}
\end{definition}
In Formula \ref{asttrans:eq:asttransDepRep}, the $ancestors(t, a)$  function returns the path from the root of $a$ to the parent node of $t$ with the root of $a$ as the first element of the list (i.e., $ancestors(t, a)[1]$). Using Definition \ref{asttrans:def:representationDepthK}, the $DepRep()$ of node $C\_1\_18$ (called node $t$) in Figure \ref{asttrans:fig:motivExample}, the sub-AST (called $a$) that have the root node as node $C\_1\_1$ in Candidate 1 and depth $k=2$ is the node $C\_1\_6$ (the node with type $binary\_expression$).  Since the depth from $t$ to the root node of the sub-AST $a$ in Candidate 1 is 5, the $DepRep(t, a, 6)$ (and the same with $k>6$) is node $C\_1\_12$, which is the parent node of $t$. 
\begin{definition}[ASTTrans Textual Representation of a Non-terminal Node]\label{asttrans:def:textRepresentationNonterminalNode}
The ASTTrans Textual Representation of a non-terminal node $n$ of AST $a$, is the mapping function from an AST non-terminal node to a string of tokens, called $TextRep(n, a)$ is defined by Formula \ref{asttrans:eq:asttransTextRep}:
\begin{equation}
    TextRep(n,a)=type(n)\#L \sum_{c \in children(n,a)}^{}type(c)\#R
    \label{asttrans:eq:asttransTextRep}
\end{equation}
Given an AST non-terminal node, this formula integrates the node type and grammatical structure into a textual representation. For example, the $TextRep$ of node $C\_1\_6$ in example shown in Figure \ref{asttrans:fig:motivExample} is the string: "$binary\_exp\#L$ $paren\_exp\#R$ $\&\&\#R$ $paren\_exp\#R$". 
\end{definition}
\begin{definition}[ASTTrans Sequence of Nodes Representation at Depth-K of an AST]\label{asttrans:def:ASTTransSeqNodeRep}
The ASTTrans Sequence of Nodes Representation at depth-k of an AST $a$ called the $NodeSeq(a,k)$, is a sequence of nodes defined by Formula \ref{asttrans:eq:asttransNodeSeq}:
\begin{equation}
    NodeSeq(a,k) = set(\sum_{l \in leaves(a)}^{}DepRep(l,a,k))
    \label{asttrans:eq:asttransNodeSeq}
\end{equation}
The $NodeSeq(a,k)$ function returns the set of nodes with max depth (from the root of the AST $a$) as $k$ for input AST $a$. Since multiple leaves can have a common ancestor node, the $set()$ function filters repetitive nodes in the output of the $NodeSeq()$ function. For the sub-AST $a$ defined in Candidate 1 in Figure \ref{asttrans:fig:motivExample}, we have $NodeSeq(a,0)$ as $\{C\_1\_1\}$, since with depth $k=0$ and sub-AST $a$, all terminal nodes of $a$ can be generated from their root node $C\_1\_1$.  The returned nodes from $NodeSeq(a,1)$ are $\{C\_1\_1, C\_1\_3, C\_1\_4\}$.   
\end{definition}
\begin{definition}[ASTTrans Textual Representation at Depth-K of an AST]\label{asttrans:def:ASTTransRepresentation}
The ASTTrans Textual Representation at depth-k of an AST $a$, called the $TextSeq(a,k)$, is a sequence of tokens generated following Formula \ref{asttrans:eq:asttransTextSeq}:
\begin{equation}
    TextSeq(a,k) =  \sum_{n \in NodeSeq(a,k)}^{}TextRep(n,a)
    \label{asttrans:eq:asttransTextSeq}
\end{equation}
Our work focuses on building models for learning ASTTrans Textual Representation at depth $k$ from an NL query.  In our standard configuration, we set $k=5$. We call this textual representation of AST \textbf{ASTTrans Representation}.
\end{definition}

%% file: content/Approach.tex
\section{Approach}
\subsection{Overview}
We propose an approach that integrates our ASTTrans Representation into the original code search process. In summary, we design a separate module of code search called the augmented code search process, which is done in parallel with the original code search process. The output as matrices of similarities generated by original code search and augmented code search processes are combined to contribute a combined matrix of similarities used to sort the code candidates by their relevancy to the given query.
\input{figTabAlgs/fig_overview_icse}

We show in detail how the augmented code search process supports the original code search process through five steps shown in the overview architecture in Figure \ref{asttrans:fig:overview}. We inherit steps 1 and 2 from the SOTA approaches \cite{002_GCB,001_UniXcoder}. In step 3, from a query and candidates, the augmented embedding as vectors for that query and candidates are generated by ASTTrans. A well-augmented embedding model requires that the vector representation of an NL query should have higher similarity to its correct candidate's vector than incorrect candidates' vectors. In step 4, the results of comparing a query's augmented embedding to each candidate's augmented embedding are calculated and logged in the similarity matrix of the augmented code search. In this problem, we use cosine similarity \cite{030_CosineSimilarityDefinition} as the metric for comparison. In steps 4 and 5, the matrix of similarity generated by augmented code search, called augmented similarity matrix, is combined with the matrix of similarity by original code search models (i.e., GraphCodeBERT and UniXcoder) and becomes a so-called Combined Similarity Matrix. Each element in this combined similarity matrix is the score of the similarity comparison between a query and each candidate. The final output of this combined code search model is the list of candidates sorted using the combined similarity matrix. While step 4 is similar to step 2 except for using different input, we discuss steps 3 and 5 below.
\subsection{Generating Augmented Embedding for Queries by Neural Machine Translation}
This module accepts the input as the query written in natural language. The expected output is the embedding as the corresponding ASTTrans representation of its respective code, i.e., the ASTTrans Representation of the correct candidate for the query. Since sequence-to-sequence translation can be solved successfully by Neural Machine Translation in prior works \cite{017_SMT_CodeMigration,018_NMT_SoftwareLocalization,027_NMT_google}, we apply NMT as a sub-module to handle this task. Two sub-modules are used for this module: NMT learning from query-to-ASTTrans Representation as a sequence of tokens and the vectorization from the sequence of tokens to vector using fastText library \cite{022_fastText}.
\input{figTabAlgs/fig_opennmt_icse}
\subsubsection{Query-to-ASTTrans Representation} The query-to-ASTTrans Representation by NMT is built in two phases. In the first phase, the training model is built to learn the mapping between natural language query and sequence of tokens as textual information from ASTTrans Representation. We build training models for datasets in the CAT benchmark \cite{004_CATBenchmark}. We inherit two advantages of learning with NMT that we illustrate in Figure \ref{asttrans:fig:augmentedEmbeddingForQuery}. First, NMT allows learning with complex textual sequences by an encoder-decoder paradigm. It includes two layers of hidden units for encoding text in the source language to embedding representation, and the other two layers decode that embedding to textual representation in the target language. While older machine translation models such as Statistical Machine Translation (SMT) \cite{017_SMT_CodeMigration} attempt to generate each sentence from phrase to phrase, NMT allows learning from longer units such as sentences or paragraphs. The NMT model considers the translation process as a continuous token generation process. Inside it, each token in the target language is generated based on the contextual information of the previous tokens in the target language and the sequence of tokens in the source language. This advantage is achieved by the attention mechanism, a module connecting the learned information from the context of source and target tokens. The output of the training phase is a trained model that can predict the ASTTrans Representation (defined in Formula \ref{asttrans:eq:asttransTextSeq}) from the input query. In the second phase, these trained models are used to predict the sequence of text as ASTTrans Representation for unseen natural language queries. We use OpenNMT \cite{003_OpenNMT} to train our models for query-to-ASTTrans Representation.
\subsubsection{Text-to-Vector Conversion.} The augmented embedding for query is completed with a module to handle the output of NMT as a sequence of tokens as the predicted ASTTrans Representation (see Figure \ref{asttrans:fig:augmentedEmbeddingForQuery}). We use fastText \cite{022_fastText} as the library for text-to-vector conversion for augmented embedding generation for each query. We train the models for vector generation of fastText with unsupervised mode. We use the ASTTrans Representation from candidates in four datasets of CAT benchmark \cite{004_CATBenchmark} to train fastText's models.

\subsection{Generating Augmented Embedding for Candidates by AST Extraction}
An important rule of code search approaches implemented by original embedding models such as GraphCodeBERT \cite{002_GCB}, and UniXcoder \cite{001_UniXcoder} is that while the query contains only information about the natural language description of the code, the candidates contains only information about the source code. Thus, while building the augmented embedding model, we must follow this rule that the augmented embedding from the query accepts only the input as a natural language description of the query while the augmented embedding of the candidates accepts the source code representations of the candidate as the input. In our design selection, while the augmented embedding of the query is the predicted ASTTrans Representation of the correct candidate of source code, the augmented embedding of a candidate is the expected ASTTrans Representation of it. 
\input{figTabAlgs/AlgmGenEmbeddingCandidate}

 Pseudocode for extracting the embedding of a candidate can be shown in Algorithm \ref{asttrans:algm:augmentEmbeddingForCandidates}. First, the AST of the candidate is generated by the $generateASTForCode()$ function. Next, all the leaves of the AST are extracted by function $extractLeaves()$. From Line 3 to Line 11, a loop through each terminal node of AST is run to extract the set of nodes that can represent the AST at depth $depthSize$ and its corresponding textual information. The $findRepresentation()$ function implements the concept of Definition \ref{asttrans:def:representationDepthK}. For each node representation extracted in Line 6, its textual information is extracted by the function $getTextRep()$, which implements Definition \ref{asttrans:def:textRepresentationNonterminalNode}. The ASTTrans textual representation of the candidate's AST at depth $depthSize$ will be transformed into a vector representation in Line 12. For text-to-vector conversion, we also use fastText \cite{022_fastText} for this task. We use the same model Similar to the vector generation for the augmented embedding of queries, augmented vectors for candidates are generated from trained model of the AST Representation of candidates in  training sets of CAT benchmark \cite{004_CATBenchmark}.

\subsection{Calculating Combined Similarity Matrix}
After generating the embedding using original embedding models (by GraphCodeBERT and UniXcoder) and using ASTTrans for a query and a list of candidates, the similarities between the vector of the query and the vector of each candidate are calculated in both the original and augmented code search process. For each pair of query-candidate, we use the cosine similarity \cite{030_CosineSimilarityDefinition} for measuring their similarity. The output of the original code search process is the original similarity matrix $matrix\_sim_{org}$, and the output of the augmented code search process is the augmented similarity matrix $matrix\_sim_{aug}$ (see Figure \ref{asttrans:fig:overview}). We combine these matrices to produce the combined similarity matrix for code search phase with ASTTrans by the following formula:
\begin{equation}
    matrix\_sim_{com}=matrix\_sim_{org}*(1-w)+matrix\_sim_{aug}*w
    \label{asttrans:eq:matrixCombination}
\end{equation}

\textbf{Selecting combined weight \textbf{w}}. In Formula \ref{asttrans:eq:matrixCombination}, the weight $w$ represents the ratio from zero to one that the code search by augmented code search process can contribute to the original code search process. We select the weight that returned the best accuracy in the augmented code search process on the validation sets of CAT benchmark \cite{004_CATBenchmark}. The selected standard weight w for matrix combination is up to \textbf{$w=0.1$}.

From the combined similarity matrix $matrix\_sim_{com}$, the list of candidates for an input query is sorted by the score between each candidate and the query. Candidates more relevant to the query appear at a higher rank than unrelated candidates.

%% file: figTabAlgs/fig_overview_icse.tex
\begin{figure}[h]
\includegraphics[width=\linewidth]{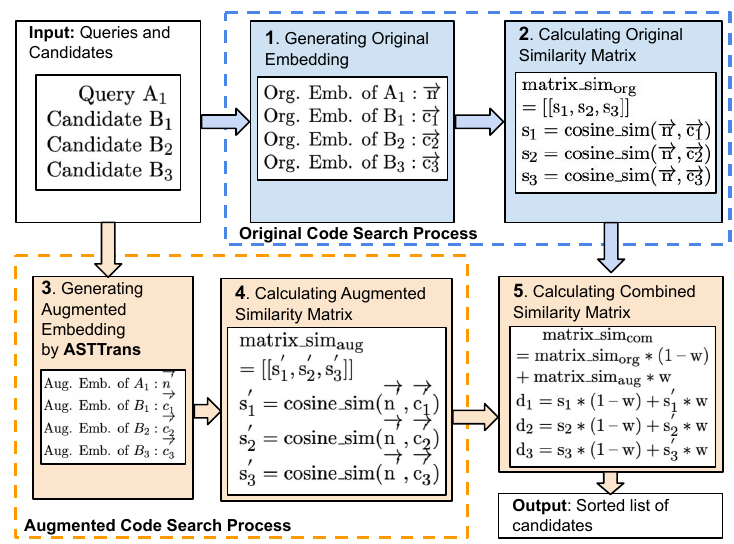}
\caption{Overview of the Combined Code Search phase from the Original/ SOTA Code Search process using GraphCodeBERT/UniXcoder and the Augmented Code Search process using ASTTrans}
\label{asttrans:fig:overview}
\end{figure}

%% file: figTabAlgs/fig_opennmt_icse.tex
\begin{figure}
\includegraphics[width=\linewidth]{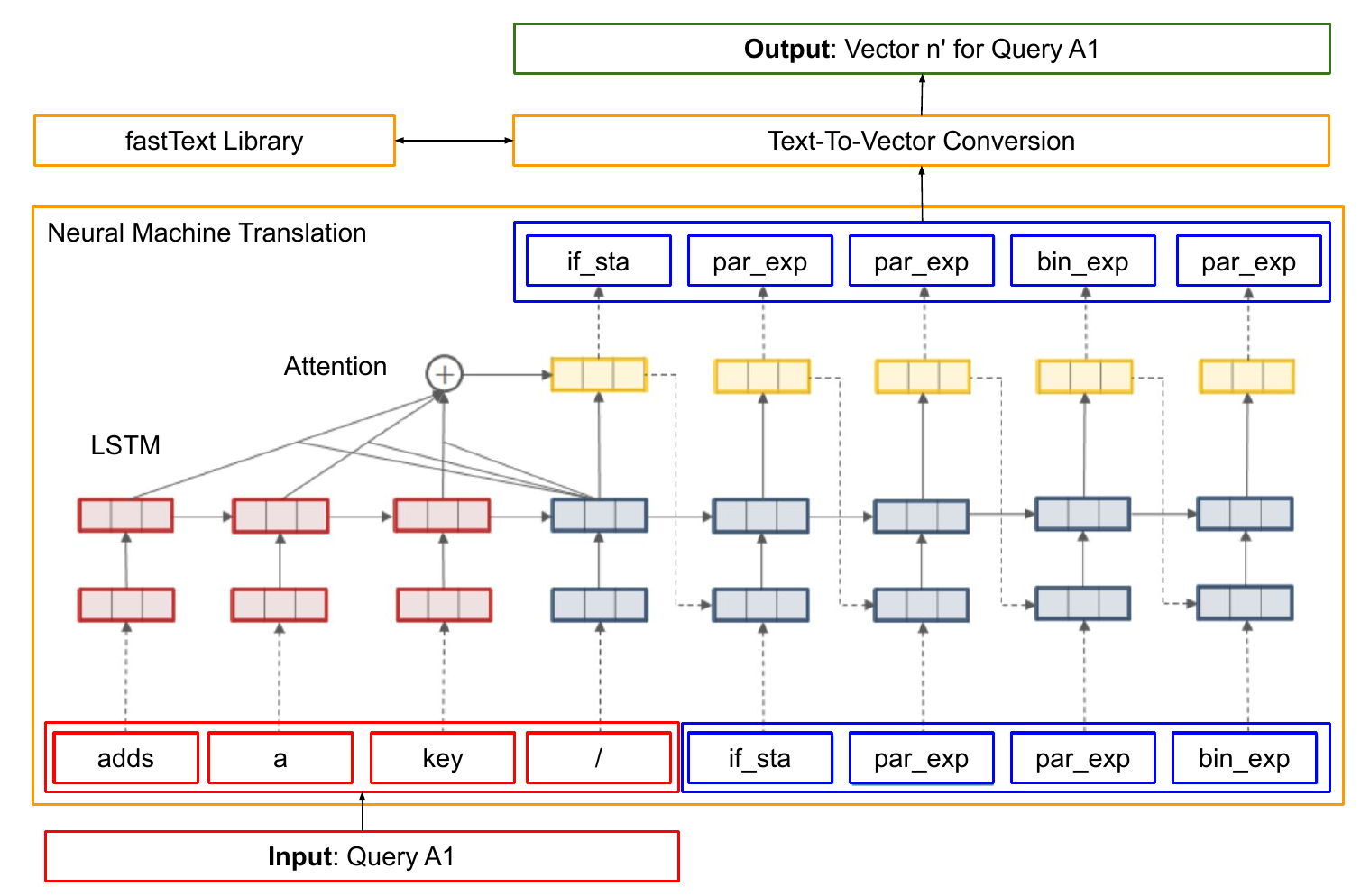}
\caption{Generating Augmented Embedding of Non-Terminal Representation of the Query's Corresponding Code Candidate by Neural Machine Translation}
\label{asttrans:fig:augmentedEmbeddingForQuery}
\end{figure}

%% file: figTabAlgs/AlgmGenEmbeddingCandidate.tex
\begin{algorithm}[h]
\SetAlgoLined
    \SetKwFunction{generateASTForCode}{generateASTForCode}
    \SetKwFunction{extractLeaves}{extractLeaves}
    \SetKwFunction{findRepresentation}{findRepresentation}
    \SetKwInOut{KwIn}{Input}
    \SetKwInOut{KwOut}{Output}
    \KwIn{candidate: string, depthSize: int, modelEmb: Model}
    \KwOut{vectorASTRep: Vector}
    
    $astCandidate\gets \generateASTForCode(candidate)$
    
    $listOfLeaves\gets \extractLeaves(astCandidate)$
    
    $seqNodeASTReps \gets set()$

    $seqTextASTReps \gets []$

    \For{$leaf  \in listOfLeaves$}{
        $nodeRep\gets \findRepresentation(leaf,depthSize)$
        
        \If{$nodeRep \notin seqNodeASTReps$}{

            $seqNodeASTReps.add(nodeRep)$

            $seqTextASTReps.append(getTextRep(nodeRep))$
         }
    }

    $vectorASTRep\gets modelEmb.getVector(seqTextASTReps)$
    
    \KwRet{$vectorASTRep$}
    \caption{Algorithm for extracting augmented embedding for code candidate }
    \label{asttrans:algm:augmentEmbeddingForCandidates}
\end{algorithm}

%% file: content/ExperimentMix.tex
\section{Experiments}
In the experiment, we attempt to answer the following research questions (RQs):
\begin{enumerate}
    \item RQ1. How well can NMT perform in learning ASTTrans Representation?
    \item RQ2. Can code search benefit from query-to-ASTTrans Representation?
    \item RQ3. How can the parameters of ASTTrans affect the performance of code search?
\end{enumerate} 
\subsection{Datasets}
\input{figTabAlgs/tbl_DatasetStat}
We use the CAT benchmark \cite{004_CATBenchmark} with four datasets of NL queries and corresponding implementation of method declarations to evaluate ASTTrans. Prior works show that code search datasets have noisy data, including erroneous code documentation/ NL queries \cite{004_CATBenchmark}. Si et al. \cite{004_CATBenchmark} proposed a systematic approach to filter noisy data. They study the four datasets to identify the templates of erroneous  parts inside each NL query. They published a clean version of them. There are three Java datasets in this benchmark, including TLCodesum (TLC) (which is the clean version of the original TLC dataset proposed in \cite{005_TLC}), Funcom (the clean version of \cite{008_Funcom}), CodeSearchNet (CSN) (the clean version of \cite{006_CSN}) and one Python dataset named PCSD (the clean version of \cite{007_PCSD}). Statistics on four datasets can be shown in Table \ref{asttrans:tbl:dataset}. 
\subsection{Configurations}
\subsubsection{OpenNMT} 
We set up the machine translation model for inferring ASTTrans Representation from the query with the following configurations. We use two sets of neural network layers for training: the encoder and decoder layers. Each module (encoder/decoder) has two layers with 500 hidden units per layer. The gate type to use in each hidden unit is Long Short Term Memory (LSTM). We choose this gate type because LSTM has been proven as an efficient Recurrent Neural Network (RNN) model that can learn and capture the relationship between words in a long textual sequence \cite{003_OpenNMT,027_NMT_google}. We use 100000 steps for training, with a validating step performed for every 1000 steps. A checkpoint will be saved for every 10000 steps. In other parameters, we use the default settings from OpenNMT \cite{003_OpenNMT}.

\subsubsection{fastText} 
We use fastText \cite{022_fastText} for text-to-vector conversion in the augmented embedding of queries and candidates. Prior work \cite{031_fastTextOutperformDoc2Vec} shows that fastText \cite{022_fastText} is not only an efficient embedding model but also able to embed a sequence of text with better quality compared to other well-known models such as Doc2Vec \cite{032_Doc2Vec} and TF-IDF \cite{033_TFIDF} in many NLP problems. In experiments, we use the augmented dimension size as 100 and the skip-gram method in fastText library for training and generating vectors for the augmented code search process.

\subsubsection{Original Embedding Models} 
We use GraphCodeBERT \cite{002_GCB} and UniXCoder \cite{001_UniXcoder} as the SOTA approaches. We use the pre-trained models for Java and Python developed by the authors of GraphCodeBERT and UniXcoder, Guo et al. \cite{002_GCB,001_UniXcoder}. We have had a few discussions with the SOTA approaches' authors about the configurations of the SOTA code search process. They confirmed that there are two settings for the code search process: without fine-tuning and with fine-tuning. The configuration without fine-tuning is called zero-short learning and always performs much less accurately than the fine-tuning setting, although the experiment without fine-tuning doesn't require a costly fine-tuning step. We select the fine-tuning setting since it reveals the best capability of SOTA approaches. We fine-tune the pre-trained models of GraphCodeBERT and UniXcoder by their proposed dataset CodeSearchNet (CSN - the full version \cite{006_CSN}). We use the fine-tuned models to generate the vectors for queries and candidates of TLC, CSN (the clean version from \cite{004_CATBenchmark}), Funcom, and PCSD datasets. Our experiments run on four datasets' test sets (see Table \ref{asttrans:tbl:dataset}).

\subsubsection{ASTTrans} 
For the standard configuration, we train four query-to-ASTTrans Representation models on four datasets in the CAT benchmark \cite{004_CATBenchmark}. There are two important parameters for ASTTrans: the depth of ASTTrans Representation and the combined weight between original and augmented similarity matrices. We set the depth $k$ of ASTTrans Representation as the \textbf{depth size $k=5$} and the \textbf{combined weight $w$ as $w=0.1$}.
\subsubsection{Embedding Size of original embedding models} 
The default dimension size (called $dim$) of a vector generated by GraphCodeBERT and UniXcoder is 768, which is much bigger than other works \cite{034_HeteroSP,032_Doc2Vec}. Reducing the dimension size by Principal Component Analysis (PCA) \cite{035_PCA} can improve running time up to five times faster with the reduced dimension size $dim$ as $dim=20$ for our code search task, as we observe in the experiments. For experiments of RQ2 and RQ3, we use two dimension sizes for dim: $dim=20$ and $dim=768$. Both original models' training and testing steps and augmented models' training and testing steps were trained on a Linux computer with 96 GB of RAM using a Core-i9 processor with 16 cores and an RTX-3080 GPU card with 24GB of RAM. 
\subsection{Metrics for Evaluation.}
\subsubsection{Evaluating Neural Machine Translation Model.} 
Prior works \cite{009_CrystalBLEU,037_METEOR_Study} show that the BLEU score, a well-known metric for evaluating NMT in NLP, has drawbacks in evaluating the quality of translated SE artifacts, such as the sequence of code tokens. Pradel et al. \cite{009_CrystalBLEU} propose an approach for filtering repetitive n-grams for calculating textual similarity for SE artifacts. They define a new metric named CrystalBLEU-4 (number four stands for four-gram cumulatively, the default configuration of CrystalBLEU). Besides, in well-known NLP metrics, Roy et al. \cite{037_METEOR_Study} show that Meteor \cite{010_Meteor} score can perform better than BLEU score by an evaluation based on human judgment. We choose \textbf{CrystalBLEU-4} and \textbf{Meteor} as metrics for evaluating the performance of NMT for RQ1.
\subsubsection{Evaluating Effect of ASTTrans in Code Search.}
Mean Reciprocal Rank (MRR) is used for code search evaluation in many approaches \cite{001_UniXcoder}. The effect in MRR over code search on set $C$ of cases by ASTTrans embedding $a$ to original embedding model $o$ with original embedding size $d$, is calculated by Formula \ref{asttrans:eq:definitionEffectMRR}:
\begin{equation}
    EffectMRR(C,a,o,d)=MRR_{com}(C,a,o,d)-MRR_{org}(C,o,d)
    \label{asttrans:eq:definitionEffectMRR}
\end{equation}
In this Formula, $MRR_{org}$ is the Original MRR returned by the original code search. $MRR_{com}$ is the Combined MRR returned by code search with the combined similarity matrix for a specific set of queries $C$, the original model $o$ reduced to dimension size $d$ by PCA \cite{035_PCA} and the augmented model $a$. If the score of this metric is positive, it means the augmented code search process improves the accuracy of the original code search process by MRR. We define the metric $Average_{EffectMRR}(C, a)$ \textbf{(Avg. Eff.)} on a set of queries $C$ using an ASTTrans model $a$ by Formula \ref{asttrans:eq:definitionAverageEffectMRR}: 
\begin{equation}
    Average_{EffectMRR}(C,a)= \frac{\sum_{d \in [20,768]}^{o \in [GCB,UniX.]}EffectMRR(C,a,o,d)}{4}
    \label{asttrans:eq:definitionAverageEffectMRR}
\end{equation}

\subsection{RQ1. How well can NMT perform in learning ASTTrans Representation? }
\input{figTabAlgs/tbl_rq1}
The result for RQ1 is shown in Table \ref{asttrans:rq1}. From the translated result of the test sets in four datasets, we see that the inference from query to code tokens confronts challenges for Neural Machine Translation. Due to the large vocabulary in the training sets, NMT achieves low accuracy measured by the CrystalBLEU-4 and Meteor scores. The best dataset for Query-to-Code Tokens translation is TLC, with the scores with a CrystalBLEU-4 score of 0.16. This dataset included over 216000 distinct tokens in their training set of 53592 candidates. Similarly, other datasets in the Java language, such as CSN and Funcom or the Python dataset PCSD, also achieve low scores by CrystalBLEU-4 and Meteor. These results confirm our assumption in practical datasets of real-world queries and source code, such as the CAT benchmark, sequences of code tokens are too complicated for the NMT model to learn their information from NL queries.

While there was low accuracy when NMT was used for Query-to-Code Tokens translation, Table \ref{asttrans:rq1} shows that Query-to-ASTTrans Representation achieves much better accuracy. Compared to the accuracy of learning code tokens, the NMT model to learn the sequence of non-terminal nodes achieves the highest accuracy in translation of the TLC dataset, while it achieves the lowest accuracy on the CSN dataset. NMT can perform more than 3x better on learning sequences of non-terminal nodes than sequences of terminal nodes as code tokens in terms of CrystalBLEU-4 score for the TLC dataset at the score of 0.51. The output of ASTTrans brings the  CrystalBLEU-4 score at 0.28 in the Funcom dataset, which is 9x better than learning code tokens. With the Python dataset PCSD, ASTTrans also achieves a significantly higher CrystalBLEU-4 score. The similarities between predicted and expected results measured by Meteor are also consistent with the first metric. One of the reasons for these improvements is that our representation of AST requires a vocabulary of types of non-terminal nodes, and its size is significantly less than the vocabulary size of learning code tokens. 
\begin{tcolorbox}
\textbf{Summary of RQ1:} NMT performs significantly better in inferring ASTTrans Representation compared to inferring code tokens from queries for CAT benchmark.
\end{tcolorbox}
\subsection{RQ2. Can code search benefit from query-to-ASTTrans Representation?}
\input{figTabAlgs/tbl_rq2_icse}
The results of RQ2 are shown in Table \ref{asttrans:tbl:rq2}. Our augmented embedding approach for queries and candidates brings positive effects (Eff.) in MRR for both test sets of four datasets. ASTTrans got the best accuracy in the TLC dataset for three Java datasets, which has the $Average_EffectMRR$ as 3.08\% for four configurations. For a low dimension size of 20, the performance using the original (Org.) embedding models decreases to 49.71\% for GraphCodeBERT and 50.12\% for UniXcoder. With the augmented code search process, the combined (Com.) code search improved by 4.47\% for GraphCodeBERT and 4.33\% for UniXCoder in MRR for this embedding size. With the high embedding size of the original model as 768, ASTTrans's augmentation improved the MRR to 1.93\% for GraphCodeBERT and 1.59\% for UniXCoder. For CSN, ASTTrans has positive effects with 0.1\% improvement in MRR on average. On the CSN dataset, ASTTrans performs best with a dimension size of 20 and embedding model as UniXcoder. ASTTrans improves the MRR to 0.15\% on average for the Funcom dataset. For this dataset, we achieve positive $EffectMRR$ scores for both four configurations using  GraphCodeBERT and UniXcoder.

ASTTrans also improves the accuracy of code search for the Python dataset PCSD with 19028 queries. It positively affects the MRR for both low-dimension and high-dimension sizes of the original embedding and for both original models. For low dimension size, ASTTrans improves 1.02\% in MRR for GraphCodeBERT and 1.45\% in MRR for UniXcoder. With a high dimension size of 768, while the original models achieve competitive accuracies of 73.38\% and 76.3\% for GraphCodeBERT and UniXcoder, ASTTrans can still improve the code search up to 0.73\% of $EffectMRR$. ASTTrans increases by 0.91\% the MRR of code search over the PCSD dataset. 
\begin{tcolorbox}
\textbf{Summary of RQ2:} ASTTrans improves the SOTA approaches with an average of 1.06\% MRR improvement in code search on CAT benchmark.    
\end{tcolorbox}

\subsection{RQ3. How can the parameters of ASTTrans affect the performance of code search?}
In this RQ, we attempt to validate the performance of ASTTrans in improving code search for SOTA embedding models with different settings of ASTTrans's parameters. We select the following parameters.
\subsubsection{Concatenating Vectors versus Combining Similarity Matrices.} 
In the default configuration, we create two matrices representing two code search processes. While the original code search process uses original embedding models, the output of the augmented code search process as the augmented similarity matrix is calculated by the augmented vectors of queries and candidates (see Figure \ref{asttrans:fig:overview}). There is another strategy to combine the embedding of ASTTrans Representation. In this alternative strategy, we concatenate elements of embedding in ASTTrans to the vectors generated by original embedding models. For example, given a query $A_{1}$, the original embedding $\overrightarrow{n}=\{n_{1},n_{2}\}$ and the augmented embedding $\overrightarrow{n^{'}}=\{n^{'}_{1},n^{'}_{2}\}$, the concatenated vector for code search is $\overrightarrow{n^{''}}=\{n_{1},n_{2},n^{'}_{1},n^{'}_{2}\}$. The concatenated vectors are then used in a single code search process. In the first part of RQ3, we change the strategy of integrating ASTTrans by using concatenated vectors and doing the code search experiment.

We show the $Average_{EffectMRR}$ of the code search with the vectors of queries and candidates by concatenated embedding in Table \ref{asttrans:tbl:rq3ConcatEmb}. The effect of MRR on four datasets significantly decreased in this configuration. The TLC, Funcom, and PCSD datasets still positively affect MRR scores in code search. Concatenated embedding strategy from ASTTrans has a negative effect on the code search on CSN dataset. The average of $Average_{EffectMRR}$ over four datasets in this configuration is 0.06\% compared to 1.06\% of the standard configuration of ASTTrans. 
\begin{tcolorbox}
\textbf{Summary of RQ3(1):} ASTTrans performs better using the combined similarity matrix than concatenated embedding.    
\end{tcolorbox}

\input{figTabAlgs/tbl_rq3_concatEmb}
\subsubsection{Combined weight between matrices.}  
\input{figTabAlgs/tbl_rq3_p2}
In the second part of this RQ, we run the code search with different weights as 0.2, 0.3, and 0.4. to analyze the performance of ASTTrans with different combined weights. We illustrate this variation by Table \ref{asttrans:tbl:rq3_weight}. We show that with a weight higher than 0.1, the augmented code search process caused negative impacts on the original model for three datasets Funcom, CSN, and PCSD. With a weight of 0.2, the augmented embedding positively impacted the TLC dataset. It shows that although ASTTrans can improve code search, integrating ASTTrans in code search require a proper approach to adjust the attention weight of their contribution to the final output of code search.
\begin{tcolorbox}
\textbf{Summary of RQ3(2):} For combined weight $w$ varies up to 0.4, ASTTrans performs best with $w=0.1$.    
\end{tcolorbox}
\subsubsection{Depth of ASTTrans Representation} 
While the default depth of ASTTrans Representation is $k=5$, we demonstrate how well a sequence of non-terminal nodes with different depths from the root can improve code search. A sequence of non-terminal nodes close to the root node of an AST might be easier for NMT to learn the information due to its simplicity. However, it has the risk of not giving enough information to distinct candidates. For example, the ASTTrans Representation at depth $k=0$ returns the root node for every candidate, meaning that representation cannot differentiate candidates for code search. Since the non-terminal nodes at $k<2$ are too abstract, we select the range of $k \in \{2-9\}$ to evaluate the code search process.
\input{figTabAlgs/fig_rq3_depth}

The average of $Average_{EffectMRR}$ result over four datasets by augmented embedding models trained from different depths for ASTTrans Representation is shown in Figure \ref{asttrans:fig:rq3_depth}. Overall, ASTTrans performs best in improving code search with our standard configuration as $k=5$. With the depth $k$ higher than 5, the improvement of $Average_{EffectMRR}$  in four datasets tends to decrease slowly. The lowest performance for code search is the configuration with  $k=2$, which supports our assumption that the set of non-terminal nodes closer to the root might be too abstract for improving code search. 
\begin{tcolorbox}
\textbf{Summary of RQ3(3):} For the depth of AST Representation $k \in \{2-9\}$, ASTTrans performs best at $k=5$ and worst at $k=2$.    
\end{tcolorbox}

%% file: figTabAlgs/tbl_DatasetStat.tex
\begin{table}
\centering
\small
\caption{Number of entities (pairs of query-candidate) for training/ validation/ test sets for four datasets of the CAT benchmark}
\begin{tabular}{|l|l|r|r|r|}
\hline
\multicolumn{1}{|c|}{\textbf{Dataset}} & \multicolumn{1}{c|}{\textbf{Language}} & \multicolumn{1}{c|}{\textbf{Training}} & \multicolumn{1}{c|}{\textbf{Validation}} & \multicolumn{1}{c|}{\textbf{Test}} \\ \hline
TLC                                    & java                                   & 53592                               & 7561                                & 7584                               \\ \hline
CSN                                    & java                                   & 323225                               & 8849                                & 19317                              \\ \hline
Funcom                                 & java                                   & 1184437                             & 20000                               & 20000                              \\ \hline
PCSD                                   & python                                 & 57846                               & 19000                               & 19028                              \\ \hline
\end{tabular}
\label{asttrans:tbl:dataset}
\end{table}

%% file: figTabAlgs/tbl_rq1.tex
\begin{table}
\small
\caption{RQ1. Comparison of NMT's performance on Query-to-ASTTrans Representation versus Query-to-Code Tokens on CAT Benchmark \cite{004_CATBenchmark}}
\begin{tabular}{|l|rrr|rrr|}
\hline
\multicolumn{1}{|c|}{\textbf{Model}}   & \multicolumn{3}{c|}{\textbf{\begin{tabular}[c]{@{}c@{}}Query-to-\\ ASTTrans Rep.\end{tabular}}}                                                      & \multicolumn{3}{c|}{\textbf{\begin{tabular}[c]{@{}c@{}}Query-to-\\ Code Tokens\end{tabular}}}                                                                        \\ \hline
\multicolumn{1}{|c|}{\textbf{Dataset}} & \multicolumn{1}{l|}{\textbf{Vocab.}} & \multicolumn{1}{c|}{\textbf{\begin{tabular}[c]{@{}c@{}}Crystal\\ BLEU-4\end{tabular}}} & \multicolumn{1}{c|}{\textbf{Meteor}} & \multicolumn{1}{l|}{\textbf{Vocab.}} & \multicolumn{1}{c|}{\textbf{\begin{tabular}[c]{@{}c@{}}Crystal\\ BLEU-4\end{tabular}}} & \multicolumn{1}{c|}{\textbf{Meteor}} \\ \hline
TLC                           & \multicolumn{1}{r|}{234}             & \multicolumn{1}{r|}{0.51}                                                              & 0.65                                 & \multicolumn{1}{r|}{216434}          & \multicolumn{1}{r|}{0.16}                                                              & 0.42                                 \\ \hline
CSN                                    & \multicolumn{1}{r|}{265}             & \multicolumn{1}{r|}{0.22}                                                              & 0.42                                 & \multicolumn{1}{r|}{596498}         & \multicolumn{1}{r|}{0.00}                                                              & 0.15                                 \\ \hline
Funcom                                 & \multicolumn{1}{r|}{274}             & \multicolumn{1}{r|}{0.28}                                                              & 0.50                                 & \multicolumn{1}{r|}{1667479}         & \multicolumn{1}{r|}{0.03}                                                              & 0.33                                 \\ \hline
PCSD                                   & \multicolumn{1}{r|}{196}             & \multicolumn{1}{r|}{0.29}                                                              & 0.56                                 & \multicolumn{1}{r|}{253253}          & \multicolumn{1}{r|}{0.04}                                                              & 0.20                                 \\ \hline
\end{tabular}
  \label{asttrans:rq1}
\end{table}

%% file: figTabAlgs/tbl_rq2_icse.tex
\begin{table}
\small
\caption{RQ2. Comparison between combined code search and original code search on CAT Benchmark \cite{004_CATBenchmark} with metrics: Combined MRR (Com.), Original MRR (Org.), Effect in MRR (Eff) and $Average_{EffectMRR}$ (Avg. Eff.)}
\begin{tabular}{|l|crrcrrc|}
\hline
                                     & \multicolumn{7}{c|}{\textbf{TLC (7584 queries)}}                                                                                                                                                                                                                                                                \\ \hline
\multicolumn{1}{|c|}{\textbf{}}      & \multicolumn{3}{c|}{\textbf{Dim.=20}}                                                                          & \multicolumn{3}{c|}{\textbf{Dim.=768}}                                                                         & \multirow{2}{*}{\textbf{\begin{tabular}[c]{@{}c@{}}Avg.\\ Eff.\end{tabular}}} \\ \cline{1-7}
\multicolumn{1}{|c|}{\textbf{Model}} & \multicolumn{1}{c|}{\textbf{Com.}} & \multicolumn{1}{c|}{\textbf{Org.}} & \multicolumn{1}{c|}{\textbf{Eff.}}   & \multicolumn{1}{c|}{\textbf{Com.}} & \multicolumn{1}{c|}{\textbf{Org.}} & \multicolumn{1}{c|}{\textbf{Eff.}}   &                                                                               \\ \hline
GCB                                  & \multicolumn{1}{r|}{54.18\%}       & \multicolumn{1}{r|}{49.71\%}       & \multicolumn{1}{r|}{\textbf{4.47\%}} & \multicolumn{1}{r|}{77.51\%}       & \multicolumn{1}{r|}{79.43\%}       & \multicolumn{1}{r|}{\textbf{1.93\%}} & \multirow{2}{*}{\textbf{3.08\%}}                                              \\ \cline{1-7}
UniX.                                & \multicolumn{1}{r|}{54.44\%}       & \multicolumn{1}{r|}{50.12\%}       & \multicolumn{1}{r|}{\textbf{4.33\%}} & \multicolumn{1}{r|}{81.32\%}       & \multicolumn{1}{r|}{79.72\%}       & \multicolumn{1}{r|}{\textbf{1.59\%}} &                                                                               \\ \hline
                                     & \multicolumn{7}{c|}{\textbf{CSN (19317 queries)}}                                                                                                                                                                                                                                                               \\ \hline
\multicolumn{1}{|c|}{\textbf{}}      & \multicolumn{3}{c|}{\textbf{Dim.=20}}                                                                          & \multicolumn{3}{c|}{\textbf{Dim.=768}}                                                                         & \multirow{2}{*}{\textbf{\begin{tabular}[c]{@{}c@{}}Avg.\\ Eff.\end{tabular}}} \\ \cline{1-7}
\multicolumn{1}{|c|}{\textbf{Model}} & \multicolumn{1}{c|}{\textbf{Com.}} & \multicolumn{1}{c|}{\textbf{Org.}} & \multicolumn{1}{c|}{\textbf{Eff.}}   & \multicolumn{1}{c|}{\textbf{Com.}} & \multicolumn{1}{c|}{\textbf{Org.}} & \multicolumn{1}{c|}{\textbf{Eff.}}   &                                                                               \\ \hline
GCB                                  & \multicolumn{1}{r|}{36.16\%}       & \multicolumn{1}{r|}{36.06\%}       & \multicolumn{1}{r|}{\textbf{0.10\%}} & \multicolumn{1}{r|}{63.45\%}       & \multicolumn{1}{r|}{63.43\%}       & \multicolumn{1}{r|}{\textbf{0.01\%}} & \multirow{2}{*}{\textbf{0.10\%}}                                              \\ \cline{1-7}
UniX.                                & \multicolumn{1}{r|}{35.11\%}       & \multicolumn{1}{r|}{34.83\%}       & \multicolumn{1}{r|}{\textbf{0.28\%}} & \multicolumn{1}{r|}{65.65\%}       & \multicolumn{1}{r|}{65.65\%}       & \multicolumn{1}{r|}{\textbf{0.00\%}} &                                                                               \\ \hline
                                     & \multicolumn{7}{c|}{\textbf{Funcom (20000 queries)}}                                                                                                                                                                                                                                                            \\ \hline
\multicolumn{1}{|c|}{\textbf{}}      & \multicolumn{3}{c|}{\textbf{Dim.=20}}                                                                          & \multicolumn{3}{c|}{\textbf{Dim.=768}}                                                                         & \multirow{2}{*}{\textbf{\begin{tabular}[c]{@{}c@{}}Avg.\\ Eff.\end{tabular}}} \\ \cline{1-7}
\multicolumn{1}{|c|}{\textbf{Model}} & \multicolumn{1}{c|}{\textbf{Com.}} & \multicolumn{1}{c|}{\textbf{Org.}} & \multicolumn{1}{c|}{\textbf{Eff.}}   & \multicolumn{1}{c|}{\textbf{Com.}} & \multicolumn{1}{c|}{\textbf{Org.}} & \multicolumn{1}{c|}{\textbf{Eff.}}   &                                                                               \\ \hline
GCB                                  & \multicolumn{1}{r|}{31.98\%}       & \multicolumn{1}{r|}{31.76\%}       & \multicolumn{1}{r|}{\textbf{0.22\%}} & \multicolumn{1}{r|}{59.27\%}       & \multicolumn{1}{r|}{59.26\%}       & \multicolumn{1}{r|}{\textbf{0.01\%}} & \multirow{2}{*}{\textbf{0.15\%}}                                              \\ \cline{1-7}
UniX.                                & \multicolumn{1}{r|}{31.70\%}       & \multicolumn{1}{r|}{31.38\%}       & \multicolumn{1}{r|}{\textbf{0.32\%}} & \multicolumn{1}{r|}{60.57\%}       & \multicolumn{1}{r|}{60.54\%}       & \multicolumn{1}{r|}{\textbf{0.03\%}} &                                                                               \\ \hline
                                     & \multicolumn{7}{c|}{\textbf{PCSD (19028 queries)}}                                                                                                                                                                                                                                                              \\ \hline
\multicolumn{1}{|c|}{\textbf{}}      & \multicolumn{3}{c|}{\textbf{Dim.=20}}                                                                          & \multicolumn{3}{c|}{\textbf{Dim.=768}}                                                                         & \multirow{2}{*}{\textbf{\begin{tabular}[c]{@{}c@{}}Avg.\\ Eff.\end{tabular}}} \\ \cline{1-7}
\multicolumn{1}{|c|}{\textbf{Model}} & \multicolumn{1}{c|}{\textbf{Com.}} & \multicolumn{1}{c|}{\textbf{Org.}} & \multicolumn{1}{c|}{\textbf{Eff.}}   & \multicolumn{1}{c|}{\textbf{Com.}} & \multicolumn{1}{c|}{\textbf{Org.}} & \multicolumn{1}{c|}{\textbf{Eff.}}   &                                                                               \\ \hline
GCB                                  & \multicolumn{1}{r|}{42.03\%}       & \multicolumn{1}{r|}{41.01\%}       & \multicolumn{1}{r|}{\textbf{1.02\%}} & \multicolumn{1}{r|}{73.80\%}       & \multicolumn{1}{r|}{73.38\%}       & \multicolumn{1}{r|}{\textbf{0.43\%}} & \multirow{2}{*}{\textbf{0.91\%}}                                              \\ \cline{1-7}
UniX.                                & \multicolumn{1}{r|}{40.03\%}       & \multicolumn{1}{r|}{38.58\%}       & \multicolumn{1}{r|}{\textbf{1.45\%}} & \multicolumn{1}{r|}{77.03\%}       & \multicolumn{1}{r|}{76.30\%}       & \multicolumn{1}{r|}{\textbf{0.73\%}} &                                                                               \\ \hline
\end{tabular}
\label{asttrans:tbl:rq2}
\end{table}

%% file: figTabAlgs/tbl_rq3_concatEmb.tex
\begin{table}
\small
\caption{RQ3 Part 1: Results by metric $Average_{EffectMRR}$ of configuration using Concatenated Embedding}
\begin{tabular}{|l|r|l|r|}
\hline
\multicolumn{1}{|c|}{\textbf{Dataset}} & \multicolumn{1}{c|}{\textbf{$Average_{EffectMRR}$}} & \multicolumn{1}{c|}{\textbf{Dataset}} & \multicolumn{1}{c|}{\textbf{$Average_{EffectMRR}$}} \\ \hline
TLC                                    & 0.16\%                                  & Funcom                                & 0.06\%                                 \\ \hline
CSN                                    & -0.03\%                                 & PCSD                                  & 0.04\%                                  \\ \hline
\end{tabular}
\label{asttrans:tbl:rq3ConcatEmb}
\end{table}

%% file: figTabAlgs/tbl_rq3_p2.tex
\begin{table}
\small
\caption{RQ3 Part 2: Results by metric $Average_{EffectMRR}$ of configurations with different combined weight $w$}
\begin{tabular}{|l|r|r|r|r|}
\hline
\multicolumn{1}{|c|}{\textbf{Weight}} & \multicolumn{1}{c|}{\textbf{w=0.1}} & \multicolumn{1}{c|}{\textbf{w=0.2}} & \multicolumn{1}{c|}{\textbf{w=0.3}} & \multicolumn{1}{c|}{\textbf{w=0.4}} \\ \hline
TLC                                   & 3.08\%                              & 2.51\%                              & -0.14\%                             & -4.54\%                             \\ \hline
CSN                                   & 0.10\%                              & -3.92\%                             & -8.80\%                             & -14.89\%                            \\ \hline
Funcom                                & 0.15\%                              & -2.31\%                             & -5.76\%                             & -10.43\%                            \\ \hline
PCSD                                  & 0.91\%                              & -1.20\%                             & -4.84\%                             & -9.63\%                             \\ \hline
\end{tabular}
\label{asttrans:tbl:rq3_weight}
\end{table}

%% file: figTabAlgs/fig_rq3_depth.tex
\begin{figure}
    \centering
    \includegraphics[width=\linewidth]{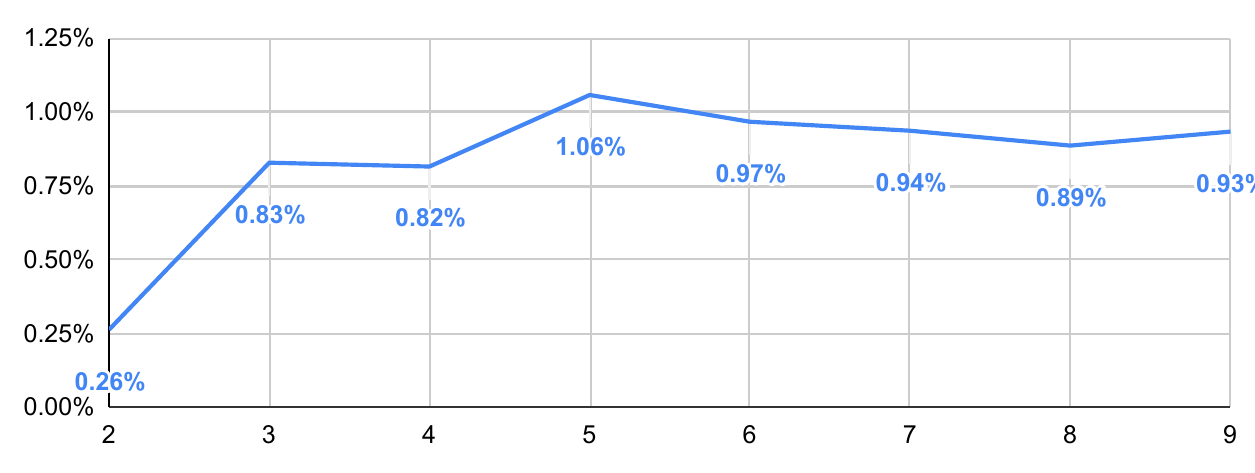}
    \caption{RQ3 Part 3: Results by average of the $Average_{EffectMRR}$ scores over four datasets with different depth $k$ of ASTTrans Representation}
    \label{asttrans:fig:rq3_depth}
\end{figure}

%% file: content/AblationStudy.tex
\section{Case Study}
\input{figTabAlgs/tbl_study_cases}

From the results in RQ2, we do a head-to-head comparison between the original results and the results from our combined code search process for each query. We call a triple of three elements: a query, a suggested list of candidates from the original code search process and a suggested list from the combined code search process as a case of search. There are three possible outputs of cases: Improved (means ASTTrans improved the rank of the correct candidate of the query), Same (the ranks are the same), and Decreased (the combined process returned better results). 

\textbf{Percentage and CC.} We measure the Cyclomatic Complexity (CC) using the Python library HFCCA \cite{038_cyclomatic_HFCCA}. We have two observations from Table \ref{asttrans:tbl:studyPercentAndCC}. First, the cases ASTTrans that can improve the original model are more than the cases ASTTrans that decrease the original accuracy in both datasets. Second, ASTTrans can improve the cases with relatively higher CC on the datasets at 3.59 compared to the average CC of all cases at 3.52. However, ASTTrans still has challenges on solving very complex code snippets, shown by the CC of Decreased cases of 3.72. 


\textbf{Cases when ASTTrans improved/ worse the results of SOTA approaches.} We found out that ASTTrans always improved the accuracy of code search when the quality of NMT on Query-to-ASTTrans Representation is high. The average CB score for the Improved cases by ASTTrans over four datasets is 0.42. The average CB score for Decreased cases is 0.15. We got the highest CB score of Improved cases for the TLCodesum dataset as 0.64 and the lowest CB score of Improved cases for the CodeSearchNet dataset as 0.25.



%% file: figTabAlgs/tbl_study_cases.tex
\begin{table}[]
\small
\caption{Percentages (Perc.) and average of CrystalBLEU-4 (CB) Cyclomatic Complexity (CC) of Improved, Decreased and all cases by ASTTrans to original models}
\begin{tabular}{|l|rrr|rrr|rr|}
\hline
                                   & \multicolumn{3}{c|}{\textbf{Improved}}                                                                        & \multicolumn{3}{c|}{\textbf{Decreased}}                                                                       & \multicolumn{2}{c|}{\textbf{All}}                                     \\ \hline
\multicolumn{1}{|c|}{\textbf{DS}}  & \multicolumn{1}{c|}{\textbf{Per.}}    & \multicolumn{1}{c|}{\textbf{CB}}   & \multicolumn{1}{c|}{\textbf{CC}} & \multicolumn{1}{c|}{\textbf{Per.}}    & \multicolumn{1}{c|}{\textbf{CB}}   & \multicolumn{1}{c|}{\textbf{CC}} & \multicolumn{1}{c|}{\textbf{CB}}   & \multicolumn{1}{c|}{\textbf{CC}} \\ \hline
TLC                                & \multicolumn{1}{r|}{26.99\%}          & \multicolumn{1}{r|}{0.64}          & 3.62                             & \multicolumn{1}{r|}{12.26\%}          & \multicolumn{1}{r|}{0.18}          & 3.79                             & \multicolumn{1}{r|}{0.50}          & 3.63                             \\ \hline
CSN                                & \multicolumn{1}{r|}{17.49\%}          & \multicolumn{1}{r|}{0.25}          & 4.18                             & \multicolumn{1}{r|}{15.84\%}          & \multicolumn{1}{r|}{0.14}          & 4.50                             & \multicolumn{1}{r|}{0.22}          & 4.05                             \\ \hline
Fun.                               & \multicolumn{1}{r|}{21.91\%}          & \multicolumn{1}{r|}{0.36}          & 1.92                             & \multicolumn{1}{r|}{16.94\%}          & \multicolumn{1}{r|}{0.17}          & 2.45                             & \multicolumn{1}{r|}{0.30}          & 2.12                             \\ \hline
PCSD                               & \multicolumn{1}{r|}{25.91\%}          & \multicolumn{1}{r|}{0.42}          & 4.65                             & \multicolumn{1}{r|}{19.23\%}          & \multicolumn{1}{r|}{0.10}          & 4.12                             & \multicolumn{1}{r|}{0.29}          & 4.30                             \\ \hline
\multicolumn{1}{|c|}{\textbf{Avg}} & \multicolumn{1}{r|}{\textbf{23.08\%}} & \multicolumn{1}{r|}{\textbf{0.42}} & \textbf{3.59}                    & \multicolumn{1}{r|}{\textbf{16.07\%}} & \multicolumn{1}{r|}{\textbf{0.15}} & \textbf{3.72}                    & \multicolumn{1}{r|}{\textbf{0.33}} & \textbf{3.52}                    \\ \hline
\end{tabular}
\label{asttrans:tbl:studyPercentAndCC}
\end{table}

%% file: content/RelatedWork.tex
\section{Related Work}
Research on code embedding models has been proposed for at least a decade \cite{039_API2Vec,040_CodeBERT,043_Code2Vec,044_CodeT5,045_PLBART,047_CodeGPT}. Before these tools, NLP models were used to represent source code API. Nguyen et al. propose API2Vec \cite{039_API2Vec}, a tool for learning the embedding of Java APIs and CSharp APIs based on Word2Vec \cite{041_Word2Vec}. Guo et al. build CodeBERT \cite{040_CodeBERT}, the pre-trained model for natural and programming languages. CodeBERT aims to strengthen its models by pre-training tasks such as Mask Language Modeling to prepare it for other downstream tasks like code search and code translation. CodeReviewer \cite{042_CodeReviewer} is a code embedding approach for representing code changes and reviews. code2vec \cite{043_Code2Vec} is another embedding tool for SE tasks such as method name generation. Other embedding models for code search are CodeT5 \cite{044_CodeT5}, which learns information of paths of AST, and PLBART \cite{045_PLBART} uses sequence-to-sequence denoising pre-training for code representation. 

Besides Neural Machine Translation, other machine translation techniques have also been applied in SE. Two of them are Statistical Machine Translation (SMT) \cite{046_SMTOriginal}, and CodeGPT \cite{047_CodeGPT}. Compared to NMT, SMT has been considered an older approach based on learning the mapping by phrases-to-phrases from the source language to the target language. SMT has been applied in code translation \cite{017_SMT_CodeMigration}, type inference \cite{016_StatType}, and behavior exception mining \cite{015_StatGen}. CodeGPT is considered the newer approach for applying translation in SE. CodeGPT provides a different mechanism for different problems of translation. 

%% file: content/ThreatsToValidity.tex
\section{Threats To Validity}
We identify several threats to the validity of our work. The threats to internal validity relate to biases and errors in experiments. Since authors of GraphCodeBERT \cite{002_GCB} and UniXcoder \cite{001_UniXcoder} didn't publish their fine-tuned models for code search, we re-trained their models on Java and Python programming languages. This could pose the threat that the SOTA models might have different performances. We asked the authors of GraphCodeBERT and UniXcoder about the validity of our configurations for fine-tuning. They confirmed that our configurations are correctly set and that our reported results of experiments on the CAT benchmark are reasonable. The threats to external validity are about the generalizability of our work. While our evaluation is only in Java and Python datasets, our algorithm for integrating ASTTrans to original code search models is language-independent and can be applied easily to other programming languages. 

%% file: content/Conclusion.tex
\section{Conclusion}
In this research, we analyze to optimize NMT by learning from the summarization of the Abstract Syntax Tree by a set of non-terminal nodes. We demonstrate that NMT can learn our AST sequence-based representation for non-terminal nodes more accurately than terminal nodes. Moreover, our proposed query-to-ASTTrans Representation model, ASTTrans, can improve the accuracy of code search by original embedding models GrraphCodeBERT \cite{002_GCB} and UniXcoder \cite{001_UniXcoder} on datasets of CAT benchmark as up to 3.08\% MRR improvement. In the future, we will further analyze the performance of other machine translation models such as CodeGPT \cite{047_CodeGPT} in learning our ASTTrans Representation and improving ASTTrans Representation for other SE tasks such as code summarization. 

%% file: content/Acknowledgement.tex
\section*{Acknowledgement}
This research was supported by the National Science Foundation under Grant number 2211982. We would also like to thank the ResearchIT team\footnote{https://researchit.las.iastate.edu/} at Iowa State University for their constant support.